\documentclass[aps,prd,twocolumn,superscriptaddress,nofootinbib,floatfix]{revtex4-2}

\usepackage{amsmath,amssymb,amsthm}
\usepackage{graphicx}
\usepackage{bm}
\usepackage[colorlinks=true,linkcolor=blue,citecolor=blue,urlcolor=blue]{hyperref}

\newtheorem{proposition}{Proposition}
\newcommand{\Kn}{\lvert K_n\rangle}
\newcommand{\dd}{\mathrm{d}}
\newcommand{\qn}[1]{[#1]_{p,q}}

\begin{document}

\title{Krylov dynamics of deformed oscillator chains:\\
localization, ballistic spreading, and indeterminacy}

\author{Tunde Joseph Osunmusanmi}
\email{Tunde.josunmusanmi@ku.ac.ae}
\affiliation{College of Computing and Mathematical Sciences, Department of Applied
Mathematics and Sciences, Khalifa University of Science and Technology, 127788 Abu Dhabi,
United Arab Emirates}

\date{\today}

\begin{abstract}
Krylov and spread complexity map quantum dynamics onto a quantum walk on a half-line whose
hopping amplitudes are the Lanczos coefficients, so that Krylov dynamics is the theory of
orthogonal polynomials and Jacobi operators. We use this correspondence to study Krylov chains
generated by the two-parameter $(p,q)$-deformed oscillator, which interpolates between the
Heisenberg--Weyl chain, the $q$-oscillator chain of double-scaled SYK, and chains with
exponentially growing Lanczos coefficients. For the deformed displacement chain the spread
complexity is the mean excitation number, its rate is the deformed momentum quadrature, and its
acceleration is $\ddot C_K=2\omega^2\langle[a,a^\dagger]\rangle$, so that the deformed commutator
acts as a force along the Krylov chain and the early-time growth is controlled by the mean
deformation. The deformation parameters separate four regimes. For $\max(p,q)<1$ the complexity
stays bounded, a localization produced by the algebra alone. For $p=1$, $q<1$ it grows
ballistically, and we obtain the exact asymptotic velocity as a theta-type series. At $p=q=1$ it
grows as $t^2$. For $\max(p,q)>1$ the Jacobi operator is not essentially self-adjoint, the
moment problem is indeterminate, and the Lanczos coefficients no longer determine the dynamics.
We make this concrete: even and odd truncations of the chain converge to two different
evolutions, which we identify as the two reflection-symmetric N-extremal solutions of the
moment problem; the odd limit carries a zero-energy atom whose weight we give in closed form;
and the one-parameter family of self-adjoint extensions yields a family of complexity curves
that coincide at early times and separate on the time scale for the Krylov wavefront to reach
infinity. Krylov complexity in this regime is defined only once a boundary condition at
infinite Krylov index is specified. The orthogonal-polynomial procedure and its verification on
the exactly solvable chains are given in the appendices.
\end{abstract}

\maketitle

\section{Introduction}
\label{sec:intro}

Quantum complexity has become a central object at the interface of quantum information,
many-body physics, and quantum gravity~\cite{Nielsen2006,StanfordSusskind2014,Susskind2016,BrownPRL2016,JeffersonMyers2017,Chapman2018}. Among the proposed
definitions, Krylov complexity of operators~\cite{Parker2019,Barbon2019,Rabinovici2021,Jian2021,DymarskySmolkin2021,CaputaDatta2021,Rabinovici2022,Hornedal2022} and
spread complexity of states~\cite{BCMW2022} are unambiguous once a generator and an initial
vector are given. The dynamics is projected onto the Krylov subspace generated by repeated
action of the Hamiltonian, the Lanczos algorithm produces an orthonormal basis in which the
Hamiltonian is tridiagonal, and the evolution becomes a quantum walk on a half-line with
on-site energies $a_n$ and hopping amplitudes $b_n$~\cite{Lanczos1950,ViswanathMuller}. Complexity is the
mean position on this chain. The framework has been applied to random-matrix, chaotic and
integrable dynamics~\cite{BMW2023,Basu2024,CMPT2024}, and comprehensive reviews are now
available~\cite{Nandy2024,Rabinovici2025}.

The program has produced sharp connections with holography. In double-scaled SYK the chord
Hamiltonian is a $q$-deformed oscillator, its Krylov chain is the chord-number chain, and spread
complexity tracks the length of the Einstein--Rosen
bridge~\cite{Berkooz2019,LinMaldacenaZhao2019,Lin2022,Rabinovici2023}. Caputa, Chen, McDonald, Sim\'on and
Strittmatter showed that the rate of growth of spread complexity of local quenches in
two-dimensional conformal field theories equals the proper radial momentum of the dual particle
in AdS$_3$~\cite{CCMSS2026}. The exactness of that result rests on an $SL(2,\mathbb{R})$
structure of the Krylov chain~\cite{CMP2022,DymarskyGorsky2020}, and the acceleration of the
complexity obeys a ``Newton's law'' fixed by the same algebra~\cite{CCMSS2026,Erdmenger2023},
in line with earlier proposals relating the growth of complexity to the momentum of infalling
particles~\cite{Susskind2020}.

Mathematically, a Krylov chain is a Jacobi matrix and Krylov dynamics is the theory of
orthogonal polynomials~\cite{ViswanathMuller,MuckYang2022,Muck2026,Murugan2026}. M\"uck and
Yang used this correspondence to work out Krylov complexity for the classical, Hahn-class and
Tricomi--Carlitz polynomials~\cite{MuckYang2022}, and Riemann--Hilbert methods have recently been
used to classify Lanczos asymptotics by the topology of the spectral measure~\cite{Murugan2026}.
Tridiagonal Jacobi-matrix methods also have a long history in quantum mechanics: the J-matrix
method of scattering~\cite{HellerYamani1974a,HellerYamani1974b,IsmailKoelink2011,IsmailKoelink2012}
and the tridiagonal representation approach
(TRA)~\cite{AlhaidariBahlouli2008,Alhaidari2017,AlhaidariBahlouli2019,Alhaidari2020} expand the
wave function in a basis in which the wave operator is tridiagonal, so that spectra, phase
shifts and wave functions follow from the associated orthogonal polynomials. Within the TRA, new
exactly solvable systems have been constructed~\cite{AlhaidariTaiwo2017,AlhaidariTaiwo2018,Taiwo2020,AlhaidariTaiwo2021},
potential functions have been engineered from prescribed spectra~\cite{AlhaidariTaiwo2023}, and
nonlinear extensions have been developed~\cite{AlhaidariTaiwo2025}. The Lanczos construction of
Krylov complexity is the same tridiagonalization, read in the time domain.
One aspect of the correspondence has remained largely unexplored: the \emph{moment problem}
itself. If the Lanczos coefficients grow fast enough, the Jacobi operator fails to be
essentially self-adjoint, the moment problem becomes indeterminate, and infinitely many spectral
measures share the same Lanczos data~\cite{Akhiezer,Chihara1978,Simon1998,Teschl2000,Schmudgen}. For local many-body
dynamics this cannot happen, because $b_n$ grows at most linearly~\cite{Parker2019}. It has been
noted that for super-linear growth the dynamics should depend on a boundary condition at the far
end of the Krylov chain, a possibility regarded so far as formal~\cite{Murugan2026}. It does
arise, however, for Krylov chains generated by deformed algebras, which are the natural algebraic
generalizations of the solvable chains above.

In this work we study Krylov chains generated by the two-parameter $(p,q)$-deformed
oscillator~\cite{ArikCoon1976,Biedenharn1989,Macfarlane1989,ChakrabartiJagannathan1991,JagannathanRao2006,BonatsosDaskaloyannis1999},
whose spectral deformation has also been used as a thermodynamic resource~\cite{Ozaydin2023}. In
earlier work on a measurement-driven quantum engine with this working medium~\cite{OT2026}, we
showed that the deformed position quadrature is essentially self-adjoint only for
$\max(p,q)\le1$, so that for larger deformation only finite-level truncations are well defined.
Here we read the same operator as a Krylov generator. Our main results are:
\begin{enumerate}
  \item For the deformed displacement chain, $C_K=\langle N\rangle$, $\dot C_K$ is the
        expectation value of the deformed momentum quadrature, and
        $\ddot C_K=2\omega^2\langle[a,a^\dagger]\rangle$ [Eq.~\eqref{eq:newton}]: the deformed
        commutator is the force on the Krylov chain. At early times
        $C_K=\omega^2t^2+\tfrac13\omega^4\delta t^4+O(t^6)$, with mean deformation
        $\delta=(p+q)/2-1$.
  \item The $(p,q)$ plane separates localized, ballistic, Hermite and indeterminate regimes
        (Fig.~\ref{fig:phase}). In the ballistic regime we derive the exact asymptotic velocity
        as a theta-type series, Eq.~\eqref{eq:velocity}.
  \item In the indeterminate regime, even and odd truncations converge to the two
        reflection-symmetric N-extremal solutions of the moment problem, the odd one carrying a
        zero-energy atom of weight~\eqref{eq:atom}. The full family of self-adjoint extensions
        gives a family of complexity curves (Fig.~\ref{fig:extensions}) that share all
        early-time data and differ at late times.
\end{enumerate}

The orthogonal-polynomial procedure we use is set out step by step in
Appendix~\ref{app:procedure}, together with two general identities for chains with vanishing
diagonal coefficients: an acceleration formula, which is a special case of the Ehrenfest
relations for spread complexity~\cite{Erdmenger2023}, and its closure criterion, which reproduces
the known classification of complexity algebras~\cite{CMP2022,Alishahiha2023,Nandy2024}.
Appendix~\ref{app:examples} verifies the procedure on the exactly solvable chains, including the
local quenches of Ref.~\cite{CCMSS2026}, and Appendix~\ref{app:numerics} summarizes the numerical
methods and checks. The main text is organized as follows. Section~\ref{sec:op} collects the
facts on Jacobi operators and moment problems that we use. Section~\ref{sec:deformed} introduces
the deformed chain, its exact identities and its regimes. Sections~\ref{sec:selfadjoint}
and~\ref{sec:indeterminate} analyze the self-adjoint and indeterminate regimes, and
Sec.~\ref{sec:discussion} discusses physical settings and open problems.

\section{Krylov chains as Jacobi operators}
\label{sec:op}

Let $H$ be a Hamiltonian and $\lvert\psi_0\rangle$ a normalized state. The Lanczos algorithm
(Appendix~\ref{app:procedure}) produces an orthonormal Krylov basis $\{\Kn\}$, with
$\lvert K_0\rangle=\lvert\psi_0\rangle$, in which
\begin{equation}
  H\Kn=a_n\Kn+b_n\lvert K_{n-1}\rangle+b_{n+1}\lvert K_{n+1}\rangle,
  \label{eq:tridiag}
\end{equation}
with $b_0=0$, $a_n\in\mathbb{R}$ and $b_n>0$. Writing
$\lvert\psi(t)\rangle=e^{-iHt}\lvert\psi_0\rangle=\sum_n\psi_n(t)\Kn$, the spread complexity is
\begin{equation}
  C_K(t)=\sum_{n\ge0}n\lvert\psi_n(t)\rvert^2=\langle\psi(t)|\mathcal{K}|\psi(t)\rangle,
  \label{eq:CK}
\end{equation}
with $\mathcal{K}=\sum_nn\,\Kn\langle K_n\rvert$. Defining $P_n(E)$ by
$\langle K_n|E\rangle=P_n(E)\langle K_0|E\rangle$, Eq.~\eqref{eq:tridiag} becomes
\begin{equation}
  EP_n=a_nP_n+b_nP_{n-1}+b_{n+1}P_{n+1},\qquad P_0=1,
  \label{eq:recurrence}
\end{equation}
so the $P_n$ are orthonormal polynomials for the spectral measure
$\dd\mu(E)=\lvert\langle K_0|E\rangle\rvert^2\dd E$, and
\begin{equation}
  \psi_n(t)=\int\dd\mu(E)\,e^{-iEt}P_n(E).
  \label{eq:psin}
\end{equation}

\textit{Moment problems.} We use the following classical
facts~\cite{Akhiezer,Simon1998,Schmudgen,Berezanskii}. If $\sum_nb_n^{-1}=\infty$ (Carleman's
criterion), the Jacobi operator $J$ defined by Eq.~\eqref{eq:tridiag} on finitely supported
vectors is essentially self-adjoint and the moment problem is determinate: the Lanczos data fix
$\mu$, and hence the Krylov dynamics. If $\sum_nb_n^{-1}<\infty$, $b_{n-1}b_{n+1}\le b_n^2$, and
$a_n=0$, Berezanskii's theorem implies that $J$ has deficiency indices $(1,1)$ and the moment
problem is indeterminate. Then $\sum_n\lvert P_n(z)\rvert^2<\infty$ for every $z\in\mathbb{C}$.
With the polynomials of the second kind $Q_n$, which obey Eq.~\eqref{eq:recurrence} for $n\ge1$
with $Q_0=0$ and $Q_1=1/b_1$, the Nevanlinna functions
\begin{align}
  B(z)&=-1+z\sum_{k\ge0}Q_k(0)P_k(z),\nonumber\\
  D(z)&=z\sum_{k\ge0}P_k(0)P_k(z)
  \label{eq:nevanlinna}
\end{align}
are entire. The self-adjoint extensions $J_\theta$ of $J$ correspond one to one to the
N-extremal solutions $\mu_\theta$ of the moment problem,
$\theta\in\mathbb{R}\cup\{\infty\}$. Each $\mu_\theta$ is discrete, supported on the zeros of
$\theta B(z)-D(z)$ (of $B$ for $\theta=\infty$), with weights
\begin{equation}
  \mu_\theta(\{x\})=\rho(x)\equiv\Bigl[\,\sum_{n\ge0}P_n(x)^2\Bigr]^{-1},
  \label{eq:rho}
\end{equation}
the largest mass that any solution can place at $x$~\cite{Akhiezer,Simon1998}. Different
extensions generate different unitary groups $e^{-iJ_\theta t}$, and therefore different Krylov
evolutions, although all share the same moments. The truncated chain $J_N$ has the Gauss
quadrature of $\mu$ as its spectral measure, with nodes at the zeros of $P_N$ and Christoffel
weights $[\sum_{n<N}P_n(x_k)^2]^{-1}$~\cite{Ismail2005}.

\section{The deformed displacement chain}
\label{sec:deformed}

\subsection{Definition and Krylov structure}

The $(p,q)$-deformed oscillator~\cite{ChakrabartiJagannathan1991} is defined by
\begin{equation}
  aa^\dagger-q\,a^\dagger a=p^N,\qquad a^\dagger a=\qn{N},
\end{equation}
with the deformed numbers
\begin{equation}
  \qn{n}=\frac{p^n-q^n}{p-q}=\sum_{r=0}^{n-1}p^{\,n-1-r}q^r,
\end{equation}
so that $[n]_{s,s}=ns^{n-1}$. The family contains the Arik--Coon oscillator
($p=1$)~\cite{ArikCoon1976} and the Biedenharn--Macfarlane oscillator
($pq=1$)~\cite{Biedenharn1989,Macfarlane1989}. We consider the displacement Hamiltonian
\begin{equation}
  H=\omega\,(a+a^\dagger),\qquad\lvert\psi_0\rangle=\lvert0\rangle,
  \label{eq:Hdef}
\end{equation}
proportional to the deformed position quadrature measured in Ref.~\cite{OT2026}. Since
$a^\dagger\lvert n\rangle=\sqrt{\qn{n+1}}\,\lvert n+1\rangle$, the Krylov basis is the Fock basis,
$\Kn=\lvert n\rangle$, with
\begin{equation}
  a_n=0,\qquad b_n=\omega\sqrt{\qn{n}} .
  \label{eq:lanczosdef}
\end{equation}
For $p=1$ and $q\to q^2$ this is the chord-number chain of double-scaled
SYK~\cite{Berkooz2019,Lin2022,Rabinovici2023}. The complexity operator is the number operator,
$\mathcal{K}=N$, so $C_K(t)=\langle N\rangle_t$, as in other settings where the Krylov and Fock
bases coincide~\cite{Adhikari2022}. For $p=q=1$ the chain reduces to the Hermite chain of
Appendix~\ref{app:examples}, $C_K=\omega^2t^2$.

\subsection{Deformed Newton's law}

The exact derivations of this subsection are given in Appendix~\ref{app:procedure}
(Lemma~1 and Propositions~\ref{prop:accel} and~\ref{prop:deformed}). For any chain,
$\dot C_K=\langle i[H,\mathcal{K}]\rangle$. Here $i[H,N]=i\omega(a-a^\dagger)$, so
\begin{equation}
  \dot C_K=i\omega\langle a-a^\dagger\rangle,
\end{equation}
which is proportional to the expectation value of the deformed momentum quadrature
$\hat\pi\propto i(a^\dagger-a)$. For chains with $a_n=0$ the acceleration obeys
\begin{equation}
  \ddot C_K=2\sum_{n\ge0}\bigl(b_{n+1}^2-b_n^2\bigr)\lvert\psi_n\rvert^2 .
  \label{eq:accel}
\end{equation}
Since $b_{n+1}^2-b_n^2=\omega^2(\qn{n+1}-\qn{n})$ and $aa^\dagger-a^\dagger a=\qn{N+1}-\qn{N}$,
\begin{equation}
  \boxed{\ddot C_K=2\omega^2\bigl\langle[a,a^\dagger]\bigr\rangle
  =2\omega^2\bigl\langle\qn{N+1}-\qn{N}\bigr\rangle .}
  \label{eq:newton}
\end{equation}
Equivalently, $[a,a^\dagger]=p^N-(1-q)\qn{N}$, Eq.~\eqref{eq:commalt}. The deformed
commutator, which measures the departure from canonical commutation of the deformed
quadratures~\cite{OT2026}, is the force on the Krylov chain. For the undeformed oscillator it
equals one, and its large-$n$ behavior in each regime is given by Eq.~\eqref{eq:forceasym}. The universal early-time expansion of Appendix~\ref{app:procedure},
Eq.~\eqref{eq:early}, with $\qn{2}=p+q$, gives
\begin{equation}
  C_K(t)=\omega^2t^2+\tfrac13\,\omega^4\delta\,t^4+O(t^6),
  \label{eq:earlydef}
\end{equation}
with $\delta=(p+q)/2-1$. The leading effect of the deformation is controlled by the mean
deformation, as for the deformed spectrum~\cite{OT2026}. The asymmetry $\gamma=(p-q)/2$ first
enters at order $t^6$ through $\qn{3}=p^2+pq+q^2$; for example, $(p,q)=(1.2,0.8)$ and $(1,1)$
both have $\delta=0$ but differ at order $t^6$. The moments of $\mu$ follow from the Dyck-path
formula~\eqref{eq:dyck}, e.g.\ $m_2=\omega^2$ and $m_4=\omega^4(1+p+q)$.

\subsection{Regimes}

\begin{figure}[t]
\centering
\includegraphics[width=0.82\columnwidth]{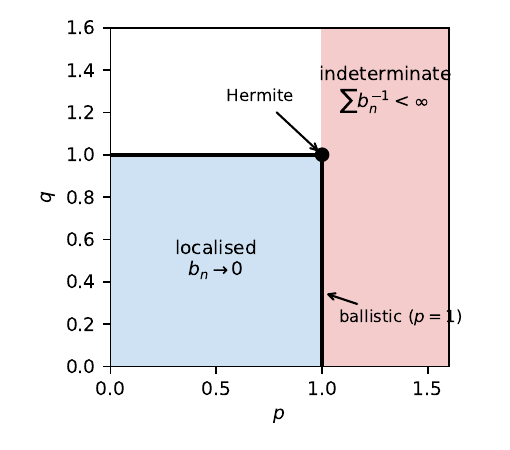}
\caption{Dynamical regimes of the deformed displacement chain, $H=\omega(a+a^\dagger)$ with
$\lvert\psi_0\rangle=\lvert0\rangle$ and Lanczos coefficients $a_n=0$, $b_n=\omega\sqrt{\qn{n}}$,
in the $(p,q)$ plane. Blue, $\max(p,q)<1$: $b_n\to0$ geometrically, $J$ is compact and $C_K$
stays bounded (localized). Thick lines, $p=1,q<1$ and $q=1,p<1$:
$b_n\to b_\infty=\omega/\sqrt{1-\min(p,q)}$ and $C_K\simeq\bar vt$ (ballistic). Black dot,
$p=q=1$: Hermite chain, $C_K=\omega^2t^2$. Red, $\max(p,q)>1$: $\sum_nb_n^{-1}<\infty$, $J$ is
not essentially self-adjoint and the moment problem is indeterminate, so the Krylov dynamics
requires a choice of self-adjoint extension.}
\label{fig:phase}
\end{figure}

The asymptotics of $\qn{n}$ separate four regimes (Fig.~\ref{fig:phase}).
\begin{itemize}
  \item $\max(p,q)<1$: $\qn{n}\to0$ geometrically, so $b_n\to0$ and $J$ is compact. The force
        $\qn{n+1}-\qn{n}$ becomes negative at large $n$.
  \item $p=1$, $q<1$ (equivalently $q=1$, $p<1$): $b_n\to b_\infty=\omega/\sqrt{1-q}$, $J$ is
        bounded, and the force decays as $q^n$.
  \item $p=q=1$: $b_n=\omega\sqrt n$ and the force is constant.
  \item $\max(p,q)>1$: $b_n\sim\max(p,q)^{n/2}$, so $\sum_nb_n^{-1}<\infty$. Since
        $\qn{n}^2-\qn{n-1}\qn{n+1}=(pq)^{n-1}>0$, also $b_{n-1}b_{n+1}\le b_n^2$, and by
        Berezanskii's theorem $J$ is not essentially self-adjoint. This is the Krylov
        counterpart of the failure of self-adjointness of the deformed quadrature found in
        Ref.~\cite{OT2026}. For $p=1$, $q>1$ the polynomials are the $q^{-1}$-Hermite
        polynomials, whose moment problem is classically
        indeterminate~\cite{Askey1989,IsmailMasson1994}.
\end{itemize}

\section{Self-adjoint regimes}
\label{sec:selfadjoint}

\begin{figure}[t]
\centering
\includegraphics[width=\columnwidth]{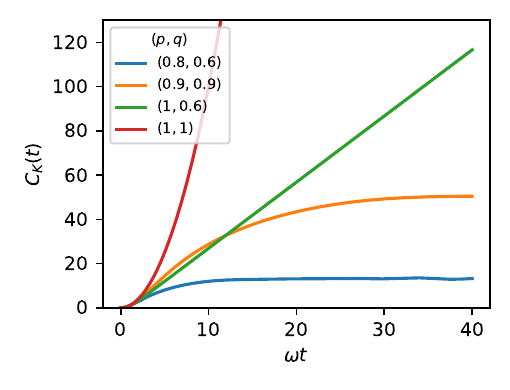}
\caption{Spread complexity $C_K(t)=\langle N\rangle_t$ of the deformed displacement chain for
four values of $(p,q)$, with $\omega=1$. Localized, $(0.8,0.6)$ and $(0.9,0.9)$: $C_K$ stays
bounded and approaches its long-time averages $13.40$ and $49.70$. Ballistic, $(1,0.6)$: linear
late-time growth with slope $\bar v=2.991\,\omega$, Eq.~\eqref{eq:velocity}. Hermite, $(1,1)$:
$C_K=t^2$ exactly. Curves were obtained by exact diagonalization of the truncated chain with
$N=800$ Krylov vectors ($N=1500$ for $p=q=1$) and are converged in $N$ to ten digits over the
time window shown.}
\label{fig:regimes}
\end{figure}

All dynamics was computed by exact diagonalization of the truncated chains $J_N$
(Appendix~\ref{app:numerics}). In the self-adjoint regimes all reported quantities are
converged in $N$ to at least ten digits. Figure~\ref{fig:regimes} shows the three behaviors.

\subsection{Localized regime}

For $\max(p,q)<1$ the operator $J$ is compact, its spectrum is pure point and accumulates only
at $E=0$, and the evolution is almost periodic. With $J\lvert E_k\rangle=E_k\lvert E_k\rangle$ and
$c_k=\langle E_k|0\rangle$, the triangle inequality gives the time-independent bound
\begin{equation}
  C_K(t)\le\Sigma^2,\qquad
  \Sigma=\sum_k\lvert c_k\rvert\,\bigl\lVert N^{1/2}\lvert E_k\rangle\bigr\rVert,
  \label{eq:bound}
\end{equation}
and the long-time average is
\begin{equation}
  \overline{C_K}=\sum_k\lvert c_k\rvert^2\langle E_k|N|E_k\rangle .
\end{equation}
For $(p,q)=(0.8,0.6)$ we find $\Sigma^2=763.85$ and $\overline{C_K}=13.3986$, both converged in
$N$, while $\max C_K$ over $0\le\omega t\le1000$ is $16.01$ and the late-time mean is $13.42$.
For $(0.9,0.9)$, $\overline{C_K}=49.699$ and $\max_{t\le1000}C_K=52.36$, with
$\Sigma^2\simeq7\times10^3$ (still slowly converging at $N=400$). The bound~\eqref{eq:bound} is
far from tight, but its finiteness establishes boundedness. Physically, the force in
Eq.~\eqref{eq:newton} is restoring beyond the site at which $\qn{n}$ is maximal (already $n=3$
for $(0.8,0.6)$), and by Eq.~\eqref{eq:forceasym} it remains negative at all larger $n$. The deformation alone thus produces a Krylov analogue of localization,
without disorder; Krylov localization has previously been associated with disorder and with
integrability~\cite{Rabinovici2022}. We conjecture that $\Sigma<\infty$ throughout $\max(p,q)<1$; a proof requires
control of the eigenvector decay near the accumulation point of the spectrum.

\subsection{Ballistic regime and exact velocity}

For $p=1$, $0<q<1$ the polynomials are the continuous $q$-Hermite polynomials. With
$E=2b_\infty\cos\theta$, the normalized spectral measure is absolutely continuous on
$[-2b_\infty,2b_\infty]$~\cite{KLS,GasperRahman2004},
\begin{equation}
  \dd\mu=\frac{(q;q)_\infty}{2\pi}\,\bigl\lvert(e^{2i\theta};q)_\infty\bigr\rvert^2\dd\theta,
  \qquad\theta\in[0,\pi].
\end{equation}

\begin{proposition}\label{prop:velocity}
For $p=1$ and $0\le q<1$, $\lim_{t\to\infty}C_K(t)/t=\bar v$, with
\begin{multline}
  \bar v=\frac{2\omega}{\pi\sqrt{1-q}}\sum_{k\in\mathbb{Z}}(-1)^kq^{k(k-1)/2}\\
  \times\left[\frac{1}{1-4k^2}-\frac{1}{1-4(k-1)^2}\right].
  \label{eq:velocity}
\end{multline}
\end{proposition}

\begin{proof}[Derivation]
Since $b_n-b_\infty=O(q^n)$, $J$ differs from the uniform half-chain $J_0$ with hopping
$b_\infty$ by a trace-class operator, so the wave operators exist and are
complete~\cite{Kato}. As $\mu$ has no point part, $e^{-iJt}\lvert0\rangle$ is asymptotic to a
free evolution $e^{-iJ_0t}\varphi$ whose spectral measure with respect to $J_0$ is $\mu$. On the
uniform chain each component with $E=2b_\infty\cos\theta$ moves with group velocity
$v=2b_\infty\sin\theta$, so $C_K/t\to\int v\,\dd\mu$. By the Jacobi triple product,
\begin{equation*}
  (q;q)_\infty\bigl\lvert(e^{2i\theta};q)_\infty\bigr\rvert^2
  =(1-e^{-2i\theta})\sum_{k\in\mathbb{Z}}(-1)^kq^{\frac{k(k-1)}{2}}e^{2ik\theta},
\end{equation*}
and $\int_0^\pi\sin\theta\,e^{2ik\theta}\dd\theta=2/(1-4k^2)$, which gives
Eq.~\eqref{eq:velocity}.
\end{proof}

\begin{figure}[t]
\centering
\includegraphics[width=\columnwidth]{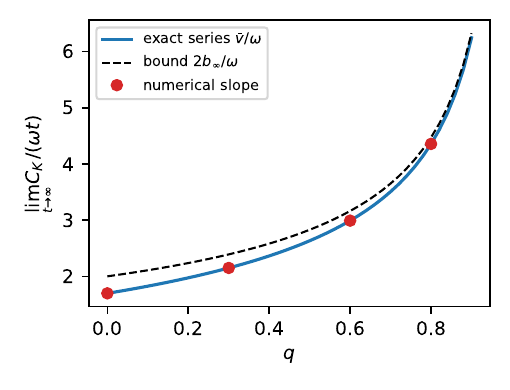}
\caption{Asymptotic spreading velocity in the ballistic regime $p=1$, $0\le q<1$. Solid: exact
series, Eq.~\eqref{eq:velocity}. Circles: late-time slope $[C_K(400)-C_K(300)]/100$ from
numerical evolution with $N=3000$ Krylov vectors, agreeing with the series to at least six
significant figures. Dashed: maximal group velocity $2b_\infty=2\omega/\sqrt{1-q}$ of the
asymptotically uniform chain, which bounds $\bar v$ because each spectral component moves with
$v=2b_\infty\sin\theta\le2b_\infty$.}
\label{fig:velocity}
\end{figure}

At $q=0$, Eq.~\eqref{eq:velocity} gives $\bar v=16\omega/(3\pi)$, the value for a uniform
half-chain started at its end. The series agrees with the late-time slope of the numerical
$C_K$ to six significant figures or better for $q=0$, $0.3$, $0.6$ and $0.8$, and to nine at
$q=0.6$ (Fig.~\ref{fig:velocity}); for instance $\bar v/\omega=2.990891$ at $q=0.6$, below the
group-velocity bound $2b_\infty/\omega=3.162$. Linear late-time growth of spread complexity in
double-scaled SYK has been discussed in Refs.~\cite{Lin2022,Rabinovici2023};
Eq.~\eqref{eq:velocity} gives its exact coefficient for the displacement chain.

\section{Indeterminate regime}
\label{sec:indeterminate}

\subsection{Parity dependence of truncations}

\begin{figure*}[t]
\centering
\includegraphics[width=0.9\textwidth]{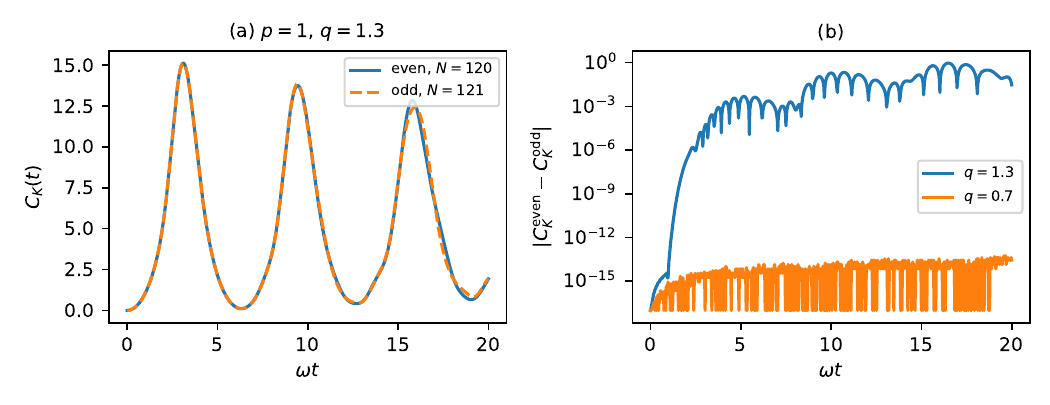}
\caption{Parity dependence of the Krylov dynamics in the indeterminate regime ($\omega=1$).
(a) $C_K(t)$ for $p=1$, $q=1.3$ from an even ($N=120$, solid) and an odd ($N=121$, dashed)
truncation. Each parity sequence is converged in $N$ (Table~\ref{tab:evenodd}); the two limits
coincide up to $\omega t\approx2.2$ and differ thereafter. (b) Absolute difference
$\lvert C_K^{\rm even}-C_K^{\rm odd}\rvert$ for $q=1.3$ (indeterminate), which grows to $0.93$,
and for $q=0.7$ (determinate, same $N$), which stays at the round-off level $10^{-13}$.}
\label{fig:evenodd}
\end{figure*}

For $\max(p,q)>1$ the Lanczos data do not fix the Krylov dynamics. We take $p=1$, $q=1.3$ and
evolve the truncated chains $J_N$ for even and odd $N$ (Table~\ref{tab:evenodd},
Fig.~\ref{fig:evenodd}). Within each parity class the results converge as $N\to\infty$, but the
two limits differ. They agree up to $\omega t\approx2.2$ to better than $10^{-6}$, as they must,
since all derivatives at $t=0$ are fixed by the common moments, and they differ by up to $0.93$
for $\omega t\le20$. At $q=0.7$, by contrast, even and odd truncations agree to $10^{-13}$. In the
indeterminate regime $C_K$ does not run away: it rises and falls quasi-periodically, consistent
with the discrete spectrum of every N-extremal solution.

\begin{table}[t]
\caption{Spread complexity $C_K(t)$ of the $p=1$, $q=1.3$ chain from even and odd truncations
($\omega=1$). The rows $\mu_\infty$ and $\mu_0$ give the N-extremal values computed without
truncation (Sec.~\ref{sec:ident}).}
\label{tab:evenodd}
\setlength{\tabcolsep}{3pt}
\begin{ruledtabular}
\begin{tabular}{rcccccc}
$N$ & $t=2$ & $t=3$ & $t=5$ & $t=6$ & $t=8$ & $t=10$\\
\hline
80  & 5.3923 & 14.9561 & 2.1385 & 0.2138 & 3.4178 & 11.3258\\
100 & 5.3925 & 14.9618 & 2.1389 & 0.2139 & 3.4174 & 11.3304\\
120 & 5.3925 & 14.9623 & 2.1389 & 0.2139 & 3.4173 & 11.3308\\
$\mu_\infty$ & 5.3925 & 14.9624 & 2.1389 & 0.2139 & 3.4173 & 11.3308\\
\hline
81  & 5.3923 & 14.9567 & 2.1414 & 0.2112 & 3.4184 & 11.1301\\
101 & 5.3925 & 14.9618 & 2.1418 & 0.2113 & 3.4180 & 11.1341\\
121 & 5.3925 & 14.9623 & 2.1418 & 0.2113 & 3.4180 & 11.1344\\
$\mu_0$ & 5.3925 & 14.9623 & 2.1418 & 0.2113 & 3.4180 & 11.1344\\
\end{tabular}
\end{ruledtabular}
\end{table}

\subsection{Identification of the limits}
\label{sec:ident}

Because $a_n=0$, $\mu$ and its truncations are reflection symmetric,
$P_n(-x)=(-1)^nP_n(x)$, and $B$ is even while $D$ is odd. Reflection therefore maps $\mu_\theta$
to $\mu_{-\theta}$, and the only symmetric N-extremal solutions are $\mu_0$ and $\mu_\infty$.
For odd $N$ the Gauss quadrature has a node at $E=0$. Since $P_{2k+1}(0)=0$ and, from
Eq.~\eqref{eq:recurrence},
\begin{equation}
  P_{2k}(0)=(-1)^k\prod_{j=1}^{k}\frac{b_{2j-1}}{b_{2j}},
\end{equation}
the weight of this node tends to $\rho(0)$:
\begin{equation}
  \boxed{\mu_0(\{0\})=\Biggl[\,\sum_{k\ge0}\,\prod_{j=1}^{k}
  \frac{\qn{2j-1}}{\qn{2j}}\Biggr]^{-1}.}
  \label{eq:atom}
\end{equation}
For $\max(p,q)\le1$ the ratios $\qn{2j-1}/\qn{2j}$ tend to one or exceed it, the series
diverges, and there is no atom. For $\max(p,q)>1$ they tend to $1/\max(p,q)$ and the series
converges. The odd limit is therefore the symmetric extension with $0$ in its spectrum,
$\mu_0$ (zeros of $D$), and the even limit is the other symmetric extension, $\mu_\infty$
(zeros of $B$).

We confirmed this directly. Evaluating Eq.~\eqref{eq:nevanlinna} in 40-digit arithmetic for
$q=1.3$, the nonnegative zeros of $D$, namely $0$, $0.96905$, $2.00518$, $3.18014$,
$4.57526,\dots$, and of $B$, namely $0.48038$, $1.47441$, $2.57051,\dots$, coincide with the
spectra of $J_{121}$ and $J_{120}$, respectively. Equation~\eqref{eq:atom} gives
$\rho(0)=0.395503$ for $q=1.3$ and $0.515793$ for $q=1.6$, in agreement with the zero-energy
weights of $J_{121}$ to six digits. Computing $C_K$ directly from the discrete measures $\mu_0$
and $\mu_\infty$, with support from the zeros of Eq.~\eqref{eq:nevanlinna} and weights from
Eq.~\eqref{eq:rho}, reproduces the truncation limits (Table~\ref{tab:evenodd}). Since a finite
atom gives a non-decaying component of the return amplitude, the long-time average of
$\lvert\psi_0(t)\rvert^2$ for the odd extension is at least $\rho(0)^2$.

\subsection{The family of extensions}

\begin{figure*}[t]
\centering
\includegraphics[width=0.9\textwidth]{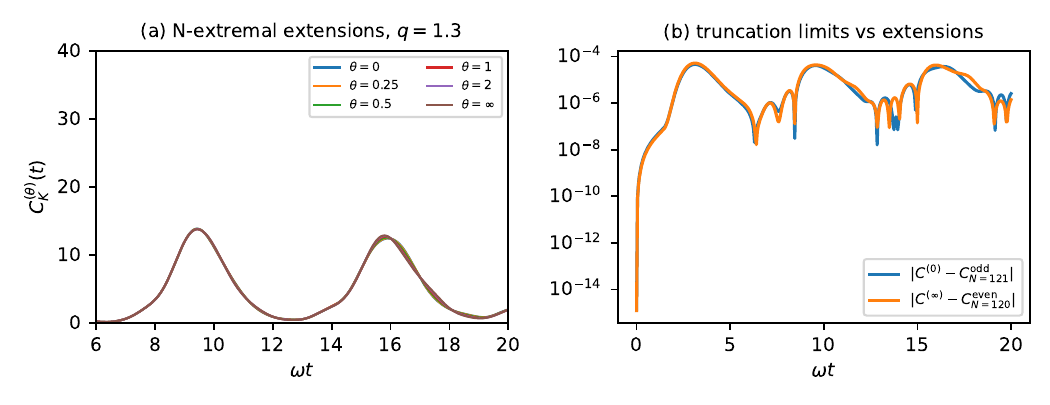}
\caption{Family of Krylov evolutions in the indeterminate regime, $p=1$, $q=1.3$, $\omega=1$.
(a) $C_K^{(\theta)}(t)$ for the N-extremal extensions $\mu_\theta$ with
$\theta=0,0.25,0.5,1,2,\infty$, computed without truncation from the zeros of $\theta B-D$,
Eq.~\eqref{eq:nevanlinna}, and the weights~\eqref{eq:rho}. The window $6\le\omega t\le20$ is
shown, where the curves separate; at earlier times they coincide, and
$C_K^{(-\theta)}=C_K^{(\theta)}$ by reflection symmetry. (b) Deviation of the truncation limits
from the two symmetric extensions: odd truncation ($N=121$) from $\theta=0$ and even truncation
($N=120$) from $\theta=\infty$. The residual $\lesssim5\times10^{-5}$ is set by the resolution of
the root search used for this figure; 40-digit evaluation at selected times agrees to seven
digits.}
\label{fig:extensions}
\end{figure*}

Every $\theta$ defines an admissible Krylov evolution. Figure~\ref{fig:extensions}(a) shows
$C_K^{(\theta)}(t)$ for several $\theta$, computed from the zeros of $\theta B-D$ and the
weights~\eqref{eq:rho}. The curves coincide at early times and separate later. At $\omega t=10$
we find $C_K^{(\theta)}=11.1344$, $11.1539$, $11.1944$, $11.2595$, $11.3063$ and $11.3308$ for
$\theta=0$, $0.25$, $0.5$, $1$, $2$ and $\infty$, and $C_K^{(\theta)}=C_K^{(-\theta)}$ by
reflection symmetry.

The separation sets in on the scale of the semiclassical time for the Krylov wavefront, moving
with $\dot n=2b_n$, to reach $n=\infty$~\cite{Murugan2026},
\begin{equation}
  T_\infty=\sum_{n\ge1}\frac{1}{2b_n}.
\end{equation}
For $p=1$ and $q=1.2$, $1.3$, $1.5$, $1.8$ and $2.2$ we find $\omega T_\infty=3.09$, $2.51$,
$1.94$, $1.56$ and $1.31$. The even and odd evolutions first differ by $10^{-3}$ at
$\omega t=10.2$, $4.0$, $2.3$, $1.6$ and $1.0$, and by $10^{-6}$ at $2.7$, $2.2$, $1.4$, $0.86$
and $0.57$. The extension acts as a boundary condition at infinite Krylov index, which the wave
function reaches in finite time and which reflects it back.

For indeterminate chains, therefore, Krylov complexity is defined only after a self-adjoint
extension, that is a boundary condition at infinite Krylov index, has been chosen. Natural
regularizations select particular extensions: odd and even truncations select $\theta=0$ and
$\theta=\infty$. All quantities fixed by finitely many moments or Lanczos coefficients are
insensitive to this choice; the late-time dynamics is not. In the engine of Ref.~\cite{OT2026},
which used $p=q=1.2$ at finite $N$, the same Jacobi operator is responsible for the absence of a
unique $N\to\infty$ limit of measurement-dependent quantities.

\section{Discussion}
\label{sec:discussion}

We have used the orthogonal-polynomial formulation of Krylov dynamics to study chains generated
by $(p,q)$-deformed oscillators. The force on the Krylov chain is the deformed commutator, which
yields localized, ballistic, Hermite and indeterminate regimes. In the ballistic regime we
obtained the exact spreading velocity. In the indeterminate regime we showed concretely that the
Lanczos data do not determine the dynamics, identified the extensions selected by natural
truncations, and computed the full family of complexity curves. For comparison, the chains of
Ref.~\cite{CCMSS2026} belong to the Meixner class, the only chains with vanishing diagonal
coefficients for which the acceleration closes on the complexity
(Proposition~\ref{prop:closure}); the deformed chains lie outside this class.

\textit{Where can the indeterminate regime arise?} Super-linear Lanczos coefficients are
excluded for local operator growth~\cite{Parker2019}, so the regime is not expected for local
operators in many-body systems; in this sense it has been regarded as
formal~\cite{Murugan2026}. It does arise in at least three settings. (i) In deformed oscillators
with $\max(p,q)>1$ used as effective models, the deformed quadrature has no unique self-adjoint
realization, and a physical model must supply one; in the measurement-driven engine of
Ref.~\cite{OT2026} this appears as a dependence of the results on the number of levels. (ii) The
same $q>1$ coordinate operator appears in the matrix-product solution of the partially
asymmetric exclusion process in its reverse-bias phase, where finite-size partition functions
are matrix elements of its powers~\cite{Derrida1993,Blythe2000,Uchiyama2004,BlytheEvans2007}. (iii) More generally, it arises for Krylov
chains of nonlocal or unbounded generators whose spectral measures decay more slowly than any
exponential~\cite{Murugan2026}. In all cases, finite-order Krylov data are universal, but the
late-time Krylov dynamics requires additional input.

Several directions remain open. (a) The Nevanlinna functions of the $q^{-1}$-Hermite problem are
known in closed form~\cite{IsmailMasson1994}, which should allow an analytic treatment of
$C_K^{(\theta)}(t)$. (b) Finiteness of $\Sigma$ in Eq.~\eqref{eq:bound} throughout the localized
regime should be proven. (c) Deformations of the $SL(2,\mathbb{R})$ chains, such as $su_q(1,1)$,
would show how the exact rate--momentum relation of Ref.~\cite{CCMSS2026} deforms when closure
fails. (d) The need for a boundary condition at infinite Krylov index, which the wave function
reaches in finite time, is reminiscent of the need for boundary conditions at the conformal
boundary of AdS, which signals also reach in finite time. Whether this analogy has content
within the holographic interpretation of spread complexity~\cite{Rabinovici2023,CCMSS2026} is an
interesting question. (e) Finally, the TRA approach to spectrum design and potential
engineering~\cite{AlhaidariTaiwo2021,AlhaidariTaiwo2023} suggests an inverse problem for Krylov
dynamics: constructing generators with prescribed Lanczos data, and hence prescribed complexity
profiles, which is a moment problem with well-developed analytic tools~\cite{Akhiezer,Ismail2005}.

\begin{acknowledgments}
Special thanks to Prof. Joan Simon of Higgs Center For Theoretical Physics at the University of Edinburgh for introducing this research topic. 
\end{acknowledgments}

\appendix

\section{Orthogonal-polynomial formulation of Krylov dynamics}
\label{app:procedure}

We collect here the procedure used throughout, in the order in which it is applied.

\textit{Step 1: Lanczos algorithm.} Starting from $\lvert K_0\rangle=\lvert\psi_0\rangle$ and
$b_0=0$, set $a_n=\langle K_n|H|K_n\rangle$ and
\begin{equation}
  \lvert A_{n+1}\rangle=(H-a_n)\Kn-b_n\lvert K_{n-1}\rangle,
\end{equation}
then $b_{n+1}=\lVert A_{n+1}\rVert$ and
$\lvert K_{n+1}\rangle=\lvert A_{n+1}\rangle/b_{n+1}$. By construction $H$ acts on the
Krylov basis as in Eq.~\eqref{eq:tridiag}, and the amplitudes obey
\begin{equation}
  i\dot\psi_n=a_n\psi_n+b_n\psi_{n-1}+b_{n+1}\psi_{n+1},
  \label{eq:dse}
\end{equation}
with $\psi_n(0)=\delta_{n0}$~\cite{ViswanathMuller,BCMW2022}.

\textit{Step 2: Polynomials and spectral measure.} Projecting Eq.~\eqref{eq:tridiag} on
generalized eigenvectors gives the recurrence~\eqref{eq:recurrence}. The $P_n$ are orthonormal
for the spectral measure $\mu$ of $\lvert\psi_0\rangle$; conversely, by Favard's theorem any
Lanczos data define such a measure~\cite{Ismail2005}.

\textit{Step 3: Amplitudes.} The Krylov amplitudes are the polynomial transforms~\eqref{eq:psin},
and the return amplitude is the characteristic function of $\mu$,
\begin{equation}
  S(t)=\langle\psi(t)|\psi_0\rangle=\psi_0(t)^*=\int\dd\mu(E)\,e^{iEt}.
\end{equation}

\textit{Step 4: Moments.} The moments
$m_k=\int E^k\dd\mu=\langle K_0|H^k|K_0\rangle=(-i)^kS^{(k)}(0)$ determine the Lanczos
coefficients through the Hankel determinants $\Delta_n=\det[m_{i+j}]_{i,j=0}^{n-1}$,
\begin{equation}
  b_n^2=\frac{\Delta_{n+1}\Delta_{n-1}}{\Delta_n^2},\qquad\Delta_0=1 .
\end{equation}
For $a_n=0$ they have the combinatorial expansion~\cite{Flajolet1980}
\begin{equation}
  m_{2k}=\sum_{\mathcal{D}_{2k}}\ \prod_{h\in\mathcal{D}_{2k}}b_h^2,\qquad m_{2k+1}=0,
  \label{eq:dyck}
\end{equation}
where the sum runs over Dyck paths of length $2k$ and the product over their up-steps, each
up-step from height $h-1$ to $h$ contributing $b_h^2$. The Stieltjes transform of $\mu$ is the
Jacobi continued fraction of the recursion method~\cite{ViswanathMuller}.

\textit{Step 5: Determinacy.} Check Carleman's criterion $\sum_nb_n^{-1}=\infty$. If it fails,
apply the theory of Sec.~\ref{sec:op}: identify the self-adjoint extensions through the
Nevanlinna functions~\eqref{eq:nevanlinna} and compute each $\mu_\theta$ from
Eq.~\eqref{eq:rho}.

\textit{Step 6: Complexity.} Evaluate $C_K=\sum_nn\lvert\psi_n\rvert^2$ either in closed form,
from the polynomial family (Appendix~\ref{app:examples}), or numerically
(Appendix~\ref{app:numerics}).

The same structure underlies the tridiagonal representation approach to exactly solvable
quantum systems~\cite{Alhaidari2017,AlhaidariBahlouli2019,Alhaidari2020} and the J-matrix method
of scattering~\cite{HellerYamani1974a,IsmailKoelink2011}; standard references on orthogonal
polynomials and Jacobi operators are Refs.~\cite{Szego1975,Chihara1978,Teschl2000,Gautschi2004,Ismail2005}. The remainder of this appendix derives the exact
identities used in the main text for chains with vanishing diagonal coefficients, $a_n=0$, and
applies them to the deformed displacement chain.

\subsection{Heisenberg equation and ladder decomposition}

For a time-independent operator $A$ and $\lvert\psi(t)\rangle=e^{-iHt}\lvert\psi_0\rangle$,
\begin{equation}
  \frac{\dd}{\dd t}\langle\psi(t)|A|\psi(t)\rangle=\langle\psi(t)|\,i(HA-AH)\,|\psi(t)\rangle
  =i\bigl\langle[H,A]\bigr\rangle ,
  \label{eq:heisenberg}
\end{equation}
since $\tfrac{\dd}{\dd t}\lvert\psi\rangle=-iH\lvert\psi\rangle$ and
$\tfrac{\dd}{\dd t}\langle\psi\rvert=i\langle\psi\rvert H$. When $a_n=0$ the Hamiltonian splits
into a lowering and a raising part,
\begin{equation}
  H=L+L^\dagger,\qquad L=\sum_{n\ge1}b_n\,\lvert K_{n-1}\rangle\langle K_n\rvert ,
  \label{eq:ladder}
\end{equation}
with $L\Kn=b_n\lvert K_{n-1}\rangle$ and $L^\dagger\Kn=b_{n+1}\lvert K_{n+1}\rangle$.

\textit{Lemma 1.} With $\mathcal{K}=\sum_nn\,\Kn\langle K_n\rvert$,
\begin{equation}
  [\mathcal{K},L]=-L,\qquad[\mathcal{K},L^\dagger]=L^\dagger,
  \label{eq:lemma1a}
\end{equation}
and $[L,L^\dagger]$ is diagonal in the Krylov basis,
\begin{equation}
  [L,L^\dagger]=\sum_{n\ge0}\bigl(b_{n+1}^2-b_n^2\bigr)\Kn\langle K_n\rvert .
  \label{eq:lemma1b}
\end{equation}

\textit{Proof.} Acting on a basis vector,
$\mathcal{K}L\Kn=(n-1)b_n\lvert K_{n-1}\rangle$ and $L\mathcal{K}\Kn=nb_n\lvert K_{n-1}\rangle$,
so $[\mathcal{K},L]\Kn=-L\Kn$ for every $n$; the second relation in
Eq.~\eqref{eq:lemma1a} is its adjoint. Next,
$LL^\dagger\Kn=b_{n+1}L\lvert K_{n+1}\rangle=b_{n+1}^2\Kn$ and
$L^\dagger L\Kn=b_nL^\dagger\lvert K_{n-1}\rangle=b_n^2\Kn$, which gives
Eq.~\eqref{eq:lemma1b}. $\square$

\subsection{Velocity and acceleration}

\begin{proposition}\label{prop:accel}
For $a_n=0$, the spread complexity obeys
\begin{align}
  \dot C_K&=i\langle L-L^\dagger\rangle
  =2\sum_{n\ge1}b_n\,\mathrm{Im}\bigl(\psi_n^*\psi_{n-1}\bigr),
  \label{eq:velocityapp}\\
  \ddot C_K&=2\bigl\langle[L,L^\dagger]\bigr\rangle
  =2\sum_{n\ge0}\bigl(b_{n+1}^2-b_n^2\bigr)\lvert\psi_n\rvert^2 .
  \label{eq:accelapp}
\end{align}
\end{proposition}

\begin{proof}
By Eq.~\eqref{eq:lemma1a}, $[H,\mathcal{K}]=-[\mathcal{K},L+L^\dagger]=L-L^\dagger$, and
Eq.~\eqref{eq:heisenberg} with $A=\mathcal{K}$ gives $\dot C_K=i\langle L-L^\dagger\rangle$.
In components, $\langle L\rangle=\sum_nb_n\psi_{n-1}^*\psi_n$ and
$\langle L^\dagger\rangle=\langle L\rangle^*$, so
$i\langle L-L^\dagger\rangle=-2\,\mathrm{Im}\langle L\rangle
=2\sum_nb_n\mathrm{Im}(\psi_n^*\psi_{n-1})$. Applying Eq.~\eqref{eq:heisenberg} to
$A=i[H,\mathcal{K}]$ gives $\ddot C_K=-\langle[H,[H,\mathcal{K}]]\rangle$. By bilinearity,
\begin{align*}
  [H,[H,\mathcal{K}]]&=[L+L^\dagger,\,L-L^\dagger]\\
  &=-[L,L^\dagger]+[L^\dagger,L]=-2[L,L^\dagger],
\end{align*}
because $[L,L]=[L^\dagger,L^\dagger]=0$. Equation~\eqref{eq:lemma1b} completes the proof.
\end{proof}

Equation~\eqref{eq:accelapp} is Eq.~\eqref{eq:accel} of the main text. It is a ``Newton's
law'' on the chain: a particle at site $n$ feels the force $F_n=2(b_{n+1}^2-b_n^2)$, outward
where the hopping increases and inward where it decreases. It is a special case of the
Ehrenfest relations for spread complexity~\cite{Erdmenger2023}. For a truncated chain $J_N$ the
same derivation holds with $b_N\equiv0$, so the last site feels the force $-2b_{N-1}^2$.

\subsection{Early-time expansion}

Iterating Eq.~\eqref{eq:dse} with $a_n=0$ gives $\psi_0=1-\tfrac12b_1^2t^2+O(t^4)$,
$\psi_1=-ib_1t+O(t^3)$ and $\psi_n=O(t^n)$, so that
$\lvert\psi_0\rvert^2=1-b_1^2t^2+O(t^4)$, $\lvert\psi_1\rvert^2=b_1^2t^2+O(t^4)$ and
$\lvert\psi_n\rvert^2=O(t^4)$ for $n\ge2$. Inserting these into Eq.~\eqref{eq:accelapp},
\begin{align*}
  \ddot C_K&=2b_1^2\lvert\psi_0\rvert^2+2(b_2^2-b_1^2)\lvert\psi_1\rvert^2+O(t^4)\\
  &=2b_1^2+2b_1^2\bigl(b_2^2-2b_1^2\bigr)t^2+O(t^4).
\end{align*}
Integrating twice with $C_K(0)=\dot C_K(0)=0$ gives
\begin{equation}
  C_K(t)=b_1^2t^2+\frac{b_1^2\,(b_2^2-2b_1^2)}{6}\,t^4+O(t^6).
  \label{eq:early}
\end{equation}
For the Hermite chain, $b_2^2=2b_1^2$ and the quartic term vanishes, consistent with
$C_K=\omega^2t^2$.

\subsection{Closure}

\begin{proposition}\label{prop:closure}
For $a_n=0$, the operator identity $-[H,[H,\mathcal{K}]]=A+B\mathcal{K}$ holds for constants
$A$, $B$ if and only if $b_n^2=c_1n+c_2n^2$, in which case $\ddot C_K=A+BC_K$.
\end{proposition}

\begin{proof}
By Proposition~\ref{prop:accel}, $-[H,[H,\mathcal{K}]]=2[L,L^\dagger]$ is diagonal with entries
$2(b_{n+1}^2-b_n^2)$, so the identity requires $2(b_{n+1}^2-b_n^2)=A+Bn$ for all $n\ge0$. Summing
from $0$ to $n-1$ with $b_0=0$ gives $b_n^2=\tfrac12An+\tfrac14Bn(n-1)$, a quadratic without
constant term; the converse is immediate.
\end{proof}

This is the $a_n=0$ form of the classification of complexity
algebras~\cite{CMP2022,Alishahiha2023,Nandy2024}. The three solutions are the Heisenberg--Weyl,
$SU(2)$ and $SL(2,\mathbb{R})$ chains of Appendix~\ref{app:examples}, the $a_n=0$ members of the
Meixner class of orthogonal polynomials.

\subsection{Application to the deformed displacement chain}

\begin{proposition}\label{prop:deformed}
For $H=\omega(a+a^\dagger)$ and $\lvert\psi_0\rangle=\lvert0\rangle$ with the
$(p,q)$-deformed oscillator, and for all $p,q>0$,
\begin{align}
  \dot C_K&=i\omega\langle a-a^\dagger\rangle=-2\omega\,\mathrm{Im}\langle a\rangle,
  \label{eq:rateapp}\\
  \ddot C_K&=2\omega^2\bigl\langle[a,a^\dagger]\bigr\rangle
  =2\omega^2\bigl\langle\qn{N+1}-\qn{N}\bigr\rangle,
  \label{eq:newtonapp}
\end{align}
and equivalently
\begin{equation}
  [a,a^\dagger]=p^N-(1-q)\qn{N}.
  \label{eq:commalt}
\end{equation}
\end{proposition}

\begin{proof}
Since $a\lvert n\rangle=\sqrt{\qn{n}}\,\lvert n-1\rangle$, the Krylov basis is the Fock basis,
$b_n=\omega\sqrt{\qn{n}}$, and the lowering operator of Eq.~\eqref{eq:ladder} is $L=\omega a$,
while $\mathcal{K}=N$. Equation~\eqref{eq:velocityapp} gives $\dot C_K=i\omega\langle
a-a^\dagger\rangle$, and since $\langle a^\dagger\rangle=\langle a\rangle^*$, this equals
$-2\omega\,\mathrm{Im}\langle a\rangle$. Equation~\eqref{eq:accelapp} gives
$\ddot C_K=2\omega^2\langle[a,a^\dagger]\rangle$. Finally, $a^\dagger a\lvert n\rangle=\qn{n}\lvert
n\rangle$ and $aa^\dagger\lvert n\rangle=\qn{n+1}\lvert n\rangle$, so
$[a,a^\dagger]=\qn{N+1}-\qn{N}$. The identity
$\qn{n+1}=q\,\qn{n}+p^n$, which follows from
$(p^{n+1}-q^{n+1})-q(p^n-q^n)=p^n(p-q)$, gives Eq.~\eqref{eq:commalt}; equivalently,
$aa^\dagger=q\,a^\dagger a+p^N$ is the defining relation of the algebra.
\end{proof}

Two features of the derivation are worth noting. First, only the lowering property
$Na=a(N-1)$ is used, never canonical commutation, so the result holds for every $(p,q)$.
Second, $i(a^\dagger-a)$ is proportional to the deformed momentum quadrature, so
Eq.~\eqref{eq:rateapp} is a deformed version of the statement that the complexity rate is a
momentum~\cite{CCMSS2026}.

\textit{Sign of the force.} Let $f_n=\qn{n+1}-\qn{n}$ and $M=\max(p,q)$. For $p\neq q$,
\begin{equation}
  f_n=\frac{p^n(p-1)-q^n(q-1)}{p-q}\simeq\frac{M^n(M-1)}{\lvert p-q\rvert}
  \quad(n\to\infty),
  \label{eq:forceasym}
\end{equation}
while on the diagonal $f_n=s^{n-1}[s-n(1-s)]$ for $p=q=s$. Hence, at large $n$, the force is
negative and exponentially small for $M<1$ (restoring, localized regime), equal to $q^n>0$ for
$p=1>q$ (decaying, ballistic regime), equal to $1$ for $p=q=1$, and positive and exponentially
growing for $M>1$ (indeterminate regime).

\subsection{Domain of validity}

The derivations above are algebraic identities and hold exactly for every finite truncation
$J_N$, with $b_N\equiv0$. For the infinite chain they hold whenever $\lvert\psi(t)\rangle$ lies in
the domains of $H$ and $\mathcal{K}$, which is the case for $\max(p,q)\le1$, where $J$ is
self-adjoint and the state has finite moments of $N$. In the indeterminate regime they hold for
each self-adjoint extension $J_\theta$, applied to states in its domain, and in particular for
the even and odd truncation limits of Sec.~\ref{sec:indeterminate}. We verified
Eqs.~\eqref{eq:velocityapp}, \eqref{eq:accelapp} and~\eqref{eq:newtonapp} against finite
differences of the numerical $C_K(t)$ at four parameter points covering all regimes
(Appendix~\ref{app:numerics}).

\section{Verified examples}
\label{app:examples}

We apply the procedure to the exactly solvable chains. Every closed form below was verified
against direct numerical evolution of the chain to $10^{-12}$.

\subsection{Heisenberg--Weyl chains}

\textit{Displacement from the vacuum.} For $H=\omega(a+a^\dagger)$ and
$\lvert\psi_0\rangle=\lvert0\rangle$, $a_n=0$ and $b_n=\omega\sqrt n$: the polynomials are Hermite
polynomials and $\mu$ is Gaussian. The state $e^{-iHt}\lvert0\rangle$ is the coherent state
$\lvert-i\omega t\rangle$, so the Krylov distribution is Poisson and $C_K=\omega^2t^2$, in
agreement with Proposition~\ref{prop:closure} ($\ddot C_K=2\omega^2$).

\textit{Coherent state under the oscillator.} For $H=\omega(N+\tfrac12)$ and
$\lvert\psi_0\rangle=\lvert\alpha\rangle$~\cite{CCMSS2026},
\begin{equation}
  a_n=\omega\bigl(n+\lvert\alpha\rvert^2+\tfrac12\bigr),\qquad b_n=\omega\lvert\alpha\rvert\sqrt n .
\end{equation}
The polynomials are Charlier polynomials and $\mu$ is Poisson on
$E_k=\omega(k+\tfrac12)$. In the displaced frame the state at time $t$ is a coherent state of
amplitude $\alpha(e^{-i\omega t}-1)$, so
\begin{equation}
  C_K(t)=4\lvert\alpha\rvert^2\sin^2(\omega t/2).
\end{equation}

\subsection{\texorpdfstring{$SU(2)$}{SU(2)} chain}

For $H=\gamma J_0+\alpha(J_++J_-)$ on spin $j$, started from $\lvert j,-j\rangle$,
\begin{equation}
  a_n=\gamma(n-j),\qquad b_n=\alpha\sqrt{n(2j-n+1)},
\end{equation}
with $n=0,\dots,2j$: a finite chain with Krawtchouk polynomials and binomial weight. With
$D_{SU}^2=4\alpha^2+\gamma^2$,
\begin{equation}
  C_K(t)=2j\,\frac{4\alpha^2}{D_{SU}^2}\,\sin^2\!\left(\frac{D_{SU}t}{2}\right),
\end{equation}
which for $\gamma=0$ reduces to $2j\sin^2(\alpha t)$, as required by
Proposition~\ref{prop:closure}.

\subsection{\texorpdfstring{$SL(2,\mathbb{R})$}{SL(2,R)} chains and local quenches}

For $H=\gamma L_0+\alpha(L_1+L_{-1})$ acting on the lowest-weight state $\lvert h\rangle$,
\begin{equation}
  a_n=\gamma(n+h),\qquad b_n=\alpha\sqrt{n(n+2h-1)},
\end{equation}
and the sign of $D^2=4\alpha^2-\gamma^2$ selects the family.

\textit{$D^2>0$: Meixner--Pollaczek.} The orthonormal Meixner--Pollaczek polynomials
$p_n(\xi)=\hat P_n^{(\lambda)}(\xi;\phi)$ satisfy~\cite{KLS}
\begin{multline}
  \xi p_n=-(n+\lambda)\cot\phi\,p_n
  +\frac{\sqrt{n(n+2\lambda-1)}}{2\sin\phi}\,p_{n-1}\\
  +\frac{\sqrt{(n+1)(n+2\lambda)}}{2\sin\phi}\,p_{n+1},
\end{multline}
with weight
\begin{equation}
  \rho_{\rm MP}(\xi)=\frac{(2\sin\phi)^{2\lambda}}{2\pi\Gamma(2\lambda)}\,
  e^{(2\phi-\pi)\xi}\,\lvert\Gamma(\lambda+i\xi)\rvert^2 .
\end{equation}
Comparison with Eq.~\eqref{eq:recurrence} gives
\begin{equation}
  \lambda=h,\quad\cos\phi=\frac{\gamma}{2\alpha},\quad\sin\phi=\frac{D}{2\alpha},\quad E=-D\xi,
\end{equation}
and $P_n(E)=(-1)^np_n(\xi)$. The beta integral
\begin{equation}
  \lvert\Gamma(\lambda+i\xi)\rvert^2=\Gamma(2\lambda)\int_{\mathbb{R}}
  \frac{e^{i\xi s}\,\dd s}{[2\cosh(s/2)]^{2\lambda}}
\end{equation}
implies, for $\lvert\mathrm{Re}\,\omega\rvert<\pi$,
\begin{equation}
  \int_{\mathbb{R}}\lvert\Gamma(\lambda+i\xi)\rvert^2e^{\omega\xi}\dd\xi
  =\frac{2\pi\Gamma(2\lambda)}{[2\cos(\omega/2)]^{2\lambda}},
\end{equation}
and hence
\begin{equation}
  \psi_0(t)=\left[\frac{\sin\phi}{\sin(\phi+iDt/2)}\right]^{2h}.
\end{equation}
The generating function of the Meixner--Pollaczek polynomials then gives
\begin{equation}
  \psi_n=\psi_0\sqrt{\frac{(2h)_n}{n!}}\,A^n,\qquad
  A=-\frac{\sinh(Dt/2)}{\sinh(Dt/2-i\phi)},
\end{equation}
a negative-binomial distribution, and
\begin{equation}
  C_K=\frac{2h\lvert A\rvert^2}{1-\lvert A\rvert^2}=\frac{2h}{\sin^2\phi}\sinh^2\frac{Dt}{2}.
\end{equation}
For the thermal local quench of Ref.~\cite{CCMSS2026}, $h=\Delta$, $D=2\pi/\beta$ and
$\phi=2\pi\epsilon/\beta$: the Meixner--Pollaczek angle is the regulator. This reproduces their
return amplitude, wave functions and spread complexity [their Eqs.~(12), (46) and (14)], the
latter as the $\epsilon\ll\beta$ limit of
\begin{equation}
  C_K=\frac{2\Delta}{\sin^2(2\pi\epsilon/\beta)}\,\sinh^2\frac{\pi t}{\beta}.
\end{equation}

\textit{$D^2<0$: Meixner.} With $\gamma=2\alpha\cosh\vartheta$,
$\lvert D\rvert=2\alpha\sinh\vartheta$ and $c=e^{-2\vartheta}$, the continuation $\phi\to
i\vartheta$ gives the Meixner recurrence, a negative-binomial weight on
$E_k=\lvert D\rvert(k+h)$, and
\begin{align}
  \psi_0(t)&=\frac{(1-c)^{2h}\,e^{-i\lvert D\rvert ht}}{(1-c\,e^{-i\lvert D\rvert t})^{2h}},
  \nonumber\\
  C_K(t)&=\frac{2h}{\sinh^2\vartheta}\,\sin^2\frac{\lvert D\rvert t}{2}.
\end{align}
For the CFT on a circle of length $L$, $\lvert D\rvert=2\pi/L$ and $\vartheta=2\pi\epsilon/L$,
and $E_k=2\pi(k+\Delta)/L$ is the tower of $L_0$ descendants.

\textit{$D=0$: Laguerre.} For $\gamma=2\alpha$ the polynomials are Laguerre polynomials with
$\nu=2h-1$, $E=\alpha x$ on $(0,\infty)$, $\psi_0=(1+i\alpha t)^{-2h}$ and
$C_K=2h\alpha^2t^2$. For the CFT on a line, $\alpha=1/(2\epsilon)$, so that
$C_K=\Delta t^2/(2\epsilon^2)$ and $\dot C_K=E_0t/\epsilon$ with $E_0=\Delta/\epsilon$, as in
Ref.~\cite{CCMSS2026}.

\section{Numerical methods and checks}
\label{app:numerics}

Spread complexities were obtained by diagonalizing the truncated Jacobi matrices $J_N$ with a
tridiagonal eigensolver and evolving exactly in the eigenbasis,
$\psi_n(t)=\sum_kV_{nk}V_{0k}e^{-iE_kt}$, so that there is no time-stepping error and the norm is
conserved to $10^{-15}$. We verified: (i) the closed forms of Appendix~\ref{app:examples} against
direct evolution, to $10^{-12}$; (ii) Eqs.~\eqref{eq:accel} and $\dot C_K=\langle
i[H,\mathcal{K}]\rangle$ against finite differences of $C_K(t)$ at four parameter points covering
all regimes, to the accuracy of the finite differences ($\sim10^{-7}$); (iii)
Eq.~\eqref{eq:earlydef} at small $t$; (iv) Eq.~\eqref{eq:velocity} against late-time slopes;
(v) Eq.~\eqref{eq:atom} against the zero-energy Christoffel weights of odd truncations; and (vi)
the identification of the truncation limits with $\mu_0$ and $\mu_\infty$, by computing the zeros
of Eq.~\eqref{eq:nevanlinna} and the weights~\eqref{eq:rho} in 40-digit arithmetic and evolving
with the resulting discrete measures. In the self-adjoint regimes results converge in $N$ to at
least ten digits; in the indeterminate regime convergence holds within each parity class
(Table~\ref{tab:evenodd}).


\end{document}